\documentclass[aps,prl,twocolumn,superscriptaddress]{revtex4}
\usepackage{makeidx}
\usepackage{graphicx}
\usepackage{amsmath}
\usepackage{amsfonts}
\usepackage{amssymb}
\usepackage{bm}
\usepackage{color}

\begin{document}
	\title{Proposed Model of the Giant Thermal Hall Effect in Two-Dimensional Superconductors: An Extension to the Superconducting Fluctuations Regime}
	\author{A. V. Kavokin}
	\affiliation{Westlake University, 18 Shilongshan Road, Hangzhou 310024, Zhejiang
		Province, China}
	\affiliation{Institute of Natural Sciences, Westlake Institute for Advanced Study,
		18 Shilongshan Road, Hangzhou 310024, Zhejiang Province, China}
	\affiliation{Spin Optics Laboratory, St. Petersburg State University, Ulyanovskaya 1, 198504 St. Petersburg, Russia}
	\author{Y. M. Galperin}
	\affiliation{A. F. Ioffe Physico-Technical Institute of Russian Academy of Sciences,  Polytekhnicheskaya 26, 194021 St. Petersburg, Russia}
	\affiliation{Department of Physics, University of Oslo, 0316 Oslo, Norway}
	\author{A. A. Varlamov}
	\affiliation{CNR-SPIN, DICII-University of Rome “Tor Vergata”, Via del Politecnico 1, 00133 Rome, Italy}
	
	
	\begin{abstract}
		We extend the thermodynamic approach for the description of the thermal
		Hall effect in two-dimensional superconductors above the critical temperature, where fluctuation Cooper pairs contribute to the conductivity, as well as in disordered normal metals where the particle-particle channel is important. We express the Hall heat conductivity
		in terms of the product of temperature derivatives of the chemical potential
		and of the magnetization of the system. Based on this general expression, we derive the analytical formalism that qualitatively reproduces the superlinear increase 
		of the thermal Hall conductivity with the decrease of temperature observed in a large variety of experimentally studied systems [Grissonnanche \textit{et al},  Nature \textbf{571}, 376 (2019)].
We also predict a non-monotonic behaviour of the thermal Hall conductivity in the regime of quantum fluctuations, in the vicinity of the second critical field and at very low temperatures. 
   
	\end{abstract}
	\flushbottom
	
	\maketitle
	
	
	The thermal Hall effect consists in a generation of a heat flow by
	a combined action of the temperature gradient $\mathbf{\nabla}T$  and magnetic field $\mathbf{H}$  perpendicular
	to it \cite{S30,MKh08,BKMMK15}. The heat current is generated in the direction that is perpendicular
	both to the magnetic field and the temperature gradient applied. This phenomenon is cognate to the Leduc-Righi effect  \cite{Abrikosov_book}, well known in metals and semiconductors, where  the temperature gradient induced in the direction $[{\mathbf{H} \times {\nabla}T}]$ is measured as a function of ${\nabla}T$. In agreement with the Wiedemann-Franz law, the thermal Hall effect in metals is usually very weak. This is clearly understandable as heat flows carried by phonons are weakly sensitive to magnetic fields ~\cite{Qin2012}.  However, recently,
	in a number of publications, a giant increase of the thermal Hall conductivity $\kappa_{yx}$ violating the Wiedemann-Franz law has been reported in several pseudogap cuprates such as La$_{1.6-x}$Nd$_{0.4}$Sr$_{x}$CuO$_{4}$, La$_{1.8-x}$Eu$_{0.2}$Sr$_{x}$CuO$_{4}$, La$_{2-x}$Sr$_{x}$CuO$_{4}$, and Bi$_{2}$Sr$_{2-x}$La$_{x}$CuO$_{6+\delta}$ \cite{Grissonnanche}. 
	
	The increase of the absolute value of $\kappa_{yx}$ by about two orders of magnitude and its negative sign in  these systems seems puzzling, at the first glance. Note, that the studied materials \cite{Grissonnanche} included both up-critical superconductors, where the conductivity is strongly influenced by fluctuation Cooper pairs, and normal metals where no Cooper pairs could be formed. While the mechanism of electric conductivity in the studied materials varied strongly, their thermal Hall conductivity demonstrated the same features, namely, the negative sign and the superlinear increase with the decrease of temperature. These discoveries have triggered the interest to comparatively large values of $\kappa_{yx}$ found also in antiferroics ~\cite{antiferroic} and the nearly ferroelectric insulator SrTiO$_{3}$~\cite{Li2019}. First experimental works were followed by a number of publications aimed at the further study and interpretation of the observed effects \cite{Dec2019, Sauls2020, Sachdev2020}. A multitude of possible reasons of the effect has been proposed for each studied system, while no unified approach to the interpretation of a giant increase of thermal Hall conductivity in up-critical superconductors and disordered metals is available till now, to the best of our knowledge.  
	
	Here we attempt at formulating a simple model that reveals the mechanism behind the observed effects and may be adapted to each particular experimental system. We develop a general thermodynamic approach that links $\kappa_{yx}$
	to the equilibrium characteristics of the systems under study. We
	consider an open circuit geometry where there is no electric current
	in the system (see Fig. 1). We shall assume that the system is in the stationary
	state that may be characterised by a constant electrochemical potential.
	This assumption will allow us to express $\kappa_{yx}$ through the
	temperature derivatives of the chemical potential and magnetization.
	Analyzing the recent experimental data on pseudo-gap cuprates
	we conclude that the giant Hall thermoconductivity found in these
	systems might take place because the temperature derivative of the magnetization shows a strong increase with the decrease of temperature, especially in the vicinity of the phase transition, while the temperature derivative of the chemical potential does not contain the smallness characteristic of the non-interacting degenerate Fermi gas   ($T/E_F$). Together, these two factors might be responsible for the giant magnitude of the effect. In this Letter, in the framework of the fluctuation theory approach, we derive the analytical expressions for $\kappa_{yx}$ both for a superconductor in various regimes and for a normal metal. These expressions  qualitatively describe the giant increase of (negative) thermal
	Hall conductivity reported in cuprates \cite{Grissonnanche}. Furthermore, we study the thermal Hall effect in the regime of quantum fluctuations: in the vicinity of the second critical magnetic field and in the limit of very low temperatures. We predict that the effect vanishes in zero temperature limit in a full agreement with the third law of thermodynamics.

	
{\it Basic definitions and the thermodynamic approach.}	To start with, let us recall that the electric and heat currents can be linked to the external electric field $\mathbf{E}$
	and temperature gradient $\nabla T$ with use of the conductivity $\hat{\sigma}(H)$,
	thermoelectric $\hat{\beta}(H)$, and heat conductivity $\hat{\kappa}(H)$ tensors as follows:
	\begin{equation}
		\begin{pmatrix}\mathbf{j}\\
			\mathbf{q}
		\end{pmatrix}=\begin{pmatrix}\hat{\sigma}\\
			\hat{\gamma}
		\end{pmatrix}\mathbf{E}-\begin{pmatrix}\hat{\beta}\\
			\hat{\kappa}
		\end{pmatrix}\nabla T, \label{flaws}
	\end{equation}
	with the Onsager relation $\hat{\gamma}(H)=-T\hat{\beta}(-H)$.
	
	The thermal Hall effect consists in the build up of the off-diagonal
	elements of $\hat{\kappa}$ in the presence of a magnetic field, as
	the scheme in Figure 1 shows.
	
	In the stationary regime, where the external circuit is broken, no
	electric current is flowing through the system, and the electrochemical
	potential of the charge carriers
	\begin{equation}
		\bar{\mu}=\mu+e^*\phi
	\end{equation}
	($\mu$ is the chemical potential, $\phi$ is the electrostatic potential, $e^*$ is the carrier's charge)
	remains constant. This statement is valid also if a temperature gradient is present in the sample. In this case, the chemical
	potential $\mu$ becomes dependent on the coordinate and, consequently, the
	internal electric field, $\bm{\mathcal{E}}$, is generated: 
	\begin{equation}
	{\mathcal{E}}_{x}=-\nabla_x\phi=-\frac{1}{e^*}\left(\frac{d\mu}{dT}\right)\nabla_{x}T.
	\end{equation}
	Under these conditions the diagonal components of the thermoelectric tensor $\hat{\beta}$ can be related to
	the temperature derivative of the chemical potential by the Kelvin formula \cite{Kelvin1854}, while the off-diagonal components of this tensor
	(arising if a magnetic field is applied) are governed by the appearance of uncompensated
 magnetization currents. 
 They can be expressed 
  in terms of the temperature
	derivative of the magnetization  (see, e.~g.,~\cite{Obraztsov64,Varlamov2013} and references therein):
	\begin{equation}
		\hat{\beta}=\begin{pmatrix}-\frac{\sigma_{xx}}{e^*}\frac{d\mu}{dT} & c\frac{dM_{z}}{dT}\\
			-c\frac{dM_{z}}{dT} & -\frac{\sigma_{yy}}{e^*}\frac{d\mu}{dT}
		\end{pmatrix}.
	\end{equation}
	Using these relations one can express the Hall  thermal flow as
	\begin{equation}
		q_{y}=-\kappa_{yx}\nabla_{x}T=\gamma_{yx}{\mathcal{E}}_{x}.
	\end{equation}
	We note that the second equality in Eq.~(5) is by no means universal. It is valid only in the open circuit geometry $(\mathbf{j}=0, \ \mathbf{q}=0)$
	in the stationary regime, where the effect of a temperature gradient can be fully accounted for by the introduction of an induced electric field (3).
	Using this substitution, one can write down the relation linking the thermal Hall conductivity
	to the temperature derivatives of the chemical potential and magnetization:
	\begin{equation}
		{\kappa}_{yx}=\frac{cT}{e^*}\left(\frac{dM}{dT}\right)\left(\frac{d\mu}{dT}\right).
		\label{kappaxy}
	\end{equation}
	One can see that the thermal Hall effect is governed by the product of the chemical
	potential and magnetization derivatives over temperature. This simple relation sheds light on the physics that is behind the recently observed giant thermal Hall effect in cuprates.
	We also note that, experimentally, the temperature gradient in $y$-direction is frequently measured rather than the thermal flow. This quantity, also known as the Righi-Leduc coefficient \cite{Abrikosov_book}, is dependent on both diagonal and non-diagonal components of the tensor $\hat{\kappa}$ and cannot be directly described by the proposed here expression for the thermodynamic contribution to ${\kappa}_{yx}$. However, we believe that the thermodynamic formula (\ref{kappaxy}) grasps the essential physics that is behind the observed effect.

	 To start with, using the obtained above general relation Eq.~(\ref{kappaxy}), we will focus on the role of fluctuating Cooper pairs in the thermal Hall effect above the superconducting phase transition. Even before doing any calculations one can expect that the effect will be huge here since the fluctuation diamagnetism, being
precursor of the Meissner effect, is giant \cite{LV09}. An additional reinforcing factor is the large value of the temperature derivative of the chemical potential of fluctuating Cooper pairs. We start from the detailed  study of the domain of the phase diagram close to the critical temperature using the Ginzburg-Landau formalism. We shall estimate the magnitude and the temperature dependence of the thermal Hall effect  in the domain of quantum fluctuations: above $H_{c2}(0)$ and at very low temperatures as well as in the high temperature limit, far above the critical temperature. Then, we shall address the thermal Hall effect in a normal metal where no Cooper pairs can be formed but the repulsive interaction in a particle-particle channel leads to the re-normalization of the electron effective mass. We show that also in this case the temperature derivative of the chemical potential is much larger than one of a degenerate Fermi gas of non-interacting electrons, which induces that superlinear increase of the thermal Hall conductivity with the decrease of temperature.
	
	\begin{figure}
	    \centering
	    \includegraphics{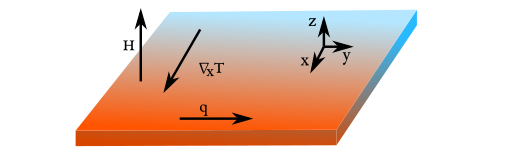}
	    \caption{The schematic showing the geometry of a thermal Hall effect measurement. The Hall bar is studied in the broken circuit geometry. The thermal flow in $y$-direction is measured as a function of the temperature gradient applied in $x$-direction and the magnetic field parallel to $z$-axis.}
	    \label{figure_1}
	\end{figure}
	
	
		{\it The free energy, magnetization, and chemical potential of fluctuating Cooper pairs.}
	We shall use the expression for the Ginzburg-Landau (GL) free energy
	for a fluctuation superconductor in the 2D case that one can find
	in Ref.~\cite{LV09}: 
	\begin{equation}
	F_{\left(2\right)}^{(\mathrm{fl})}(\epsilon,h)\!=-\frac{T_{c0}S}{4\pi\xi^{2}}\left[\epsilon\ln\frac{1}{2h}\!-\!2h\ln\frac{\Gamma(\frac{1}{2}+\frac{\epsilon}{2h})}{\sqrt{2\pi}}\right].\label{gfe2}
	\end{equation}
	Here ${S}$ is the sample cross-section and $\xi^2= \pi \mathcal {D}/8T_{c0}$ is the superconducting coherence
	length,  $\mathcal{D}$ is the electron diffusion coefficient,  $\epsilon=\ln T/T_{c0} \approx (T-T_{c0})/T_{c0} \ll 1$ is the reduced temperature,
	$T_{c0}$ is critical temperature of the superconducting phase transition
	at zero magnetic field. The dimensionless magnetic field $h={H}/{\widetilde{H}_{c2}(0)} \ll 1$
	is normalized with the second critical field $\widetilde{H}_{c2}(0)=\Phi_{0}/2\pi\xi^{2}$,
	introduced as the linear extrapolation to zero temperature of the GL formula and $\Phi_{0}=  \pi c/e$ as the magnetic flux quantum.
	Note that superconducting fluctuations behave as 2D objects since the
 characteristic size of the fluctuating Cooper pairs, $\xi(\epsilon)=\xi/\sqrt{\epsilon}$, exceeds  the thickness  $d$  of the film.
	
	The expression for $2D$ fluctuation magnetization per unit square
	of the film can be obtained just differentiating the expression for the free
	energy over magnetic field and taking this derivative with the opposite
	sign \cite{MP00}:
	\begin{eqnarray}
	M_{\left(2\right)}^{(\mathrm{fl})}(\epsilon,h)&=&\frac{T_{c0}}{\Phi_{0}}\left\{ \ln\frac{\Gamma(\frac{1}{2}+\frac{\epsilon}{2h})}{\sqrt{2\pi}}\!\right.\nonumber \\ &&
	\left.-\frac{\epsilon}{2h}\left[\psi(\frac{1}{2}+\frac{\epsilon}{2h})-1\right]\!\right\} ,\label{m2d}
	\end{eqnarray}
	where $\psi(z)=d\ln{\Gamma(x)}/dx$ is the logarithmic derivative
	of the Euler gamma function. This formula describes the crossover
	from the weak field linear regime to the saturation of the fluctuation
	magnetization in strong fields \cite{KLB73}. The temperature derivative
	of the fluctuation magnetization is given by 
	\begin{eqnarray}
	\frac{d M_{\left(2\right)}^{(\mathrm{fl})}(\epsilon,h)}{d T}  &=&
	-\frac{1}{2h\Phi_{0}}\left[ \frac{\epsilon}{2h}\psi'\left(\frac{1}{2}+\frac{\epsilon}{2h}\right)-1\right]\\
	& = & \frac{h}{\Phi_{0}}\left\{ \begin{array}{rl}
	1/6\epsilon^2, 
	& h\ll\epsilon\ll1,\nonumber\\
	1/2h^2,
	& \epsilon\ll h\ll1,\nonumber\\
	1/\epsilon_h^2 
	, &
	\epsilon_h \ll h ,
	\end{array}\right.\label{limits}
	\end{eqnarray}
	where $\epsilon_h \equiv \epsilon +h$.
	
	
	The last ingredient which we need in order 
	to be able to calculate Eq.~(\ref{kappaxy}) explicitly is the chemical potential of fluctuating Cooper pairs.
	We are interested in the expression that would be valid for an arbitrary
	relation between the temperature and the  magnetic field both in the vicinity
	of $T_{c0}$ and at high temperatures, far from $T_{c0}$. It can be  found from the definition of the chemical
	potential in terms of the derivative of the free energy, see Eq.~(\ref{gfe2}),
	over the fluctuation Cooper pairs concentration $N_{(2)}^{(\mathrm{fl})}$. The latter can be
	easily obtained by means of integration of the distribution function of the Cooper pairs over momenta. This procedure yields: 
	\begin{eqnarray}
	\mu_{\left(2\right)}^{(\mathrm{fl})}(\epsilon,h)& = & 
	-T_{\mathrm{c0}}\epsilon\,\frac{\ln\frac{1}{2h}-\frac{2h}{\epsilon}\ln\frac{\Gamma(1/2+\epsilon/2h)}{\sqrt{2\pi}}}{\ln\frac{1}{2h}-\psi\left(\frac{1}{2}+\frac{\epsilon}{2h}\right)}.\label{mu2d}
	\end{eqnarray}
	The details of this derivation are given in the Supplemental material \cite{Supplemental}.

{\it Thermal Hall conductivity due to fluctuating Cooper pairs close to $T_{\mathrm{c0}}$.}	
	Now, the thermal
	Hall conductivity can be represented explicitly: 
		\begin{eqnarray} \label{kappaxy1}
	&&	\tilde{\kappa}_{yx\left(2\right)}^{(\mathrm{fl})}(\epsilon,h)=-\frac{T_{c0}}{4 \pi h} \left[1-
\frac{\epsilon}{2h}\psi'\left(\frac{1}{2}+\frac{\epsilon}{2h}\right)\right]
\label{GLk} \\ && \quad \times
\left[ 1-\frac{\epsilon}{2h}\psi'\left(\frac{1}{2}+\frac{\epsilon}{2h}\right) \frac{\ln\frac{1}{2h}-\frac{2h}{\epsilon}\ln\frac{\Gamma(1/2+\epsilon/2h)}{\sqrt{2\pi}}}{\left[\ln\frac{1}{2h}-\psi\left(\frac{1}{2}+\frac{\epsilon}{2h}\right)\right]^{2}}\right]. \nonumber
	\end{eqnarray}
	It is instructive to express Eq.~(\ref{kappaxy1}) in its asymptotic form:
	\begin{equation}
	\tilde{\kappa}_{yx\left(2\right)}^{(\mathrm{fl})}(\epsilon,h)=-\frac{e \mathcal{D}H}{64c}\left\{ \begin{array}{rl}
	1/3\epsilon^{2}, & h\ll\epsilon\ll1,\\
	1/h^{2}, & \epsilon\ll h\ll1,\\
	\epsilon^{-2}_h\ln \frac{2h}{\epsilon_h}, & \epsilon_h\ll h .
	\end{array}\right.\label{limits1}
	\end{equation}
	 Here we operate with the true magnetic field, $H=h\widetilde{H}_{c2}(0)$. We took advantage of the relation between the GL extrapolation of the second critical field and the BCS one: $\widetilde{H}_{c2}(0)=(8 \gamma_E/\pi^2){H}^{\text{BCS}}_{c2}(0)$.  We note that in the BCS theory $ H_{c2}(0) = (2/ \gamma_E )\Phi_{0} \left(T_{c0}/\mathcal{D}\right) $, where $\gamma_E= 1.78$ is the Euler constant. 
	
	\begin{figure}
	    \centering
	    \includegraphics[width=8.6 cm]{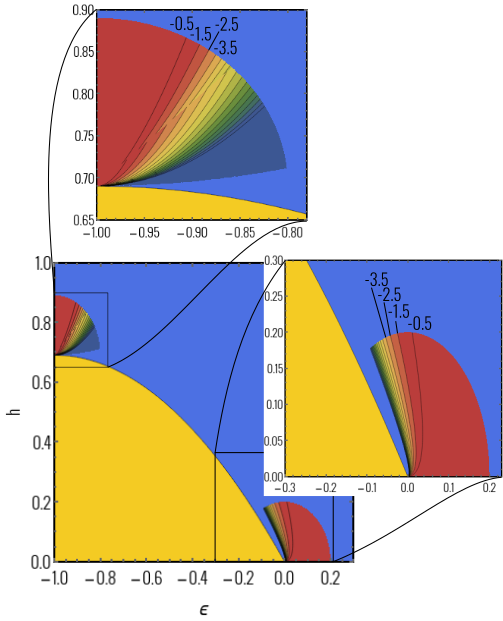}
	    \caption{Fluctuation induced thermal Hall conductivity $\tilde{\kappa}_{yx\left(2\right)}^{(\mathrm{fl})}$ (shown by the color scale and numbers in arbitrary units) as a function of dimensionless temperature and magnetic field. The blue area shows the normal phase above the superconducting transition, the yellow area  corresponds to the superconducting state. Our consideration is valid in the blue area in the domain close to the critical temperature $T_{c0}$ (see Eq. (\ref{GLk})) and in the vicinity of the second critical  field $H_{c2}(0)$ (see Eq. (\ref{qk})). }
	    \label{figure_2}
	\end{figure}
	
	
		{\it Fluctuation superconductor: the high temperature limit $T \gg T_{c0}$.} In this limit, fluctuation Cooper pairs can still be formed, and their concentration can be estimated based on the fluctuation theory~\cite{Supplemental}. Assuming that the Cooper pairs obey the Bose-Einstein statistics, we obtain a relation between the chemical potential and the concentration of Cooper pairs that allows to obtain the temperature derivative of the chemical potential. Expressing the magnetization as in Refs~\cite{AL74, Bulaevski}, we obtain the thermal Hall conductivity as:
	\begin{eqnarray*}
\kappa_{(2)}^{\text{(fl)}}\left(\!T\!\gg \!T_{c0}\right)\!=\!- \frac{2\pi^{2}}{3}\frac{eDH}{c}\!
\left[\!1-\!\frac{\ln\left(\ln\ln\frac{1}{T_{\mathrm{c0}}\tau}\!-\!\ln\ln\frac{T}{T_{c0}}\right)}{\ln (T/T_{c0})}\!\right]\! \!.           \label{High_T}
\end{eqnarray*}
	Here $\tau$ is a characteristic scattering time. We conclude that the temperature dependence of the thermal Hall conductivity becomes less pronounced as one gets farther from the phase transition point, however, the superlinear character of this dependence and the negative sign of $\kappa_{yx}$ are maintained. 
	
	{\it Normal metal. Contribution to the thermal Hall effect due to electron-electron interactions.}
		We consider a normal metal, where no fluctuation Cooper pairs is formed but the repulsive Coulomb interaction between electrons is important. This is the regime that is observed in some of the cuprates where no superconducting phase transition is observed. We analyse the temperature dependencies of the magnetization and the chemical potential here. It is important to emphasize the role of interelectron interaction in the particle-particle channel that may be considered as a counterpart of Cooper pairing and leads to the re-normalisation of the electron effective mass \cite{ARV83}. Due to this re-normalization that is strongly temperature dependent, the derivative of the chemical potential over temperature strongly varies as a function of temperature, which is crucial for the understanding of the temperature dependence of the thermal Hall conductivity. As shown in the Supplemental material, the following expression for $\kappa_{yx}$ is valid in this case:
		\begin{equation}
		\kappa_{(2)}^{(g)}=-\frac{\pi}{3}\frac{eDH}{c}\frac{1}{T\tau\ln^{2}(T_{K}/T)}.\  \label{Metal}\end{equation}
	Note, that the sign of the thermal Hall conductivity is negative, and it increases with the temperature decrease in a qualitative similarity to the behaviour characteristic of up-critical superconductors.
	
	{\it Effect of quantum fluctuations on the thermal Hall conductivity above $H_{c2}(0)$.}
	Using the general thermodynamic relation (\ref{kappaxy}) one can predict the behavior of thermal Hall conductivity above $H_{c2}(0)$ also in the limit of very low temperatures, in the domain of quantum fluctuations. The behaviour of the  fluctuation magnetization in this regime was studied in Ref. \cite{GL01}:
	\begin{equation}
	M_{\left(2\right)}^{(\mathrm{fl})} (t,\widetilde{h}) =\frac{T_{c0}}{\gamma_E\Phi_0}
	\left[\ln \frac{1}{2\mathit{\gamma }_{E}t}-\frac{\mathit{\gamma
		}_{E}t}{\widetilde{h}}-\psi \left( \frac{\widetilde{h}}{2\mathit{\gamma }_{E}t}\right) \!\right] \label{M0}
	\end{equation}
	with  $t=T/T_{c0} \ll 1$ and $\widetilde{h}=\left( H-H_{c2}(T)\right) /H_{c2}(T) \ll 1$. 
	The differentiation of Eq.~(\ref{M0}) results in
	\begin{eqnarray} \label{18}
	\frac{dM_{\left(2\right)}^{(\mathrm{fl})} (t,\widetilde{h})}{dT} &=&  
	 \frac{1}{\gamma_{E}\Phi_0}  \left[\frac{\widetilde{h}}{2\mathit{\gamma }_{E}}\frac{
		}{t^2}\psi' \left( \frac{1}{2\mathit{\gamma }_{E}}\frac{
		\widetilde{h}}{t}\right) - \frac{1}{t}-\frac{\mathit{\gamma
		}_{E}}{\widetilde{h}}\right] \nonumber\\
	&=&\frac{1}{\Phi_0} \left\{ \begin{array}{rl}
	2\gamma_{E}t/3\widetilde{h}^2, & t \ll\widetilde{h} \ll 1, \\
	1/\widetilde{h}, & \widetilde{h} \ll t \ll 1. \\
	\end{array}\right.
	\end{eqnarray}
	In the vicinity of $H_{c2}(0)$, the chemical potential of fluctuation Cooper pairs can be written as $ \mu^{ \mathrm{(QF)}} = -\Delta_{\mathrm{BCS}}\,\widetilde{h}$ (similarly to the expression valid at $T_{\mathrm{c0}}$, see Ref. \cite{VGG18}. 
	Its temperature derivative differs from zero due to the temperature dependence of $H_{c2}(T)$ (see Ref. \cite{Sarma}):
	\begin{equation}
	\frac{d\mu^{(\mathrm{QF})}}{dT} =\frac{\Delta_{\mathrm{BCS}} }{H_{c2}(0)}\left(\frac{dH_{c2}(T)}{dT}\right) =-\frac{2 \gamma_E}{\pi}t\,.
	\label{muQF}
	\end{equation}
	Substituting Eqs.~(\ref{18}) and (\ref{muQF}) into Eq.~(\ref{kappaxy}) and taking into account that $T_{c0}=(\pi/\gamma_E)\Delta_{\mathrm{BCS}}$ one finally finds 
	\begin{eqnarray}
	\tilde{\kappa}_{yx\left(2\right)}^{(\mathrm{fl})}(t,\widetilde{h}) &\! =\! &
\frac{ \Delta_{\mathrm{BCS}}}{\pi} \left[t\!+\!\frac{\mathit{\gamma
		}_{E} t^2}{\widetilde{h}}\!-\!\frac{\widetilde{h}}{2\mathit{\gamma }_{E}}
	\psi'\! \left(\! \frac{1}{2\mathit{\gamma }_{E}}\frac{
		\widetilde{h}}{t}\right)\!\right]	\label{qk} \\
	& = & -\frac{ \Delta_{\mathrm{BCS}}}{\pi} t^2\left\{ \begin{array}{rl}
	2\gamma_{E}t/3\widetilde{h}^2 , & t \ll\widetilde{h} \ll 1,\nonumber \\
1/\widetilde{h}, & \widetilde{h} \ll t \ll 1.\nonumber\\
	\end{array}\right.
	\end{eqnarray}	
	One can see that the thermal Hall conductivity vanishes at zero temperature,  in a full agreement with the third law of thermodynamics.

	{\it Results and discussion.}
	Figure 2 shows the thermal Hall conductivity as a function of the reduced
	temperature and magnetic field with zooms to the areas corresponding to the low magnetic field and quantum fluctuation regimes. One can see that, while the specific shape of the dependence may vary, $\kappa_{yx}^{(\mathrm{fl})}(\epsilon,h)$
	always has a negative sign, and its absolute value increases rapidly with the temperature decrease. The same universal behaviour is found in the high temperature limit and in normal metals, as discussed above.
	
	Now one can compare the predictions of our 
	theory with the experimental results reported in Ref.~\cite{Grissonnanche}
	for four cuprates. 
	One can notice that the theory correctly
	reproduces both the sign of the thermal conductivity and the dramatic increase of its magnitude with the temperature decrease. We believe that the qualitative agreement of such a simple model with the large variety of experimental results is significant as it hints at the essentially  thermodynamic nature of the giant thermal Hall effect.
	
	We note that the  approach we used for the description of up-critical superconductors is based  on the conventional theory of fluctuations~\cite{LV09} applicable to superconductors above the phase transition boundary $H_{c2}(T)$. It may not account for all the specifics of the experimentally studies cuprate superconductors. Yet, it turns out that the main ingredients required for application of Eq.~(\ref{kappaxy}), i.~e., temperature dependencies of the fluctuation magnetization and chemical potential of the preformed Cooper pairs in the pseudogap state, qualitatively do not differ much from that ones of a conventional superconductor. This is confirmed in the recent study Ref.~\cite{BCVL18}, that went beyond the weak-fluctuation formalism, applied the precursor-pairing approach within the BCS-BEC crossover scheme~\cite{ CSTL99, CW14} and found a large singular diamagnetic response for the temperatures much higher than the transition temperature side by side with the strong temperature dependence of the pair chemical potential in a striking similarity to the effects predicted by the simple model developed here.
	

	
	We thank P.S. Grigoryev for helping us with numerical calculations. A.K. acknowledges Project No. 041020100118 and Program 2018R01002 supported
	by Leading Innovative and Entrepreneur Team Introduction Program of
	Zhejiang. A.K. also acknowledges the support from the St-Petersburg State university within the Grant ID 51125686. A.A.V. acknowledges a support by  EU Horizon 2020  research and innovation program  under the grant agreement n 731976 (MAGENTA).

\end{document}